\documentclass[prb,draft,twocolumn,superscriptaddress,floatfix]{revtex4}
\usepackage{amsmath,amssymb}
\usepackage[final]{graphicx}
\usepackage{version}

\newcommand{\bk}{{\bf k}}

\newcommand{\br}{{\bf r}}

\newcommand{\bq}{{\bf q}}

\newcommand{\beqa}{\begin{eqnarray}}
\newcommand{\eeqa}{\end{eqnarray}}

\begin{document}


\title{``Dilute'' excitons in a double layer system: single-exciton and 
mean-field approach}
\author{Christopher R Jamell}
\email[]{crjamell@iupui.edu}
\affiliation{Department of Physics, Indiana University-Purdue University
Indianapolis (IUPUI), Indianapolis, Indiana, USA 46202}
\author{Chang-hua Zhang}
\affiliation{Department of Physics, Indiana University-Purdue University
Indianapolis (IUPUI), Indianapolis, Indiana, USA 46202}
\affiliation{Department of Physics, Kansas State University, Manhattan, 
Kansas, USA 66506}
\author{Yogesh N. Joglekar}
\affiliation{Department of Physics, Indiana University-Purdue University
Indianapolis (IUPUI), Indianapolis, Indiana, USA 46202}
\date{\today}

\begin{abstract}
Double layer systems where one layer has electrons and the holes are in a 
parallel layer a distance $d$ away are expected to undergo excitonic 
condensation 
at low temperature. This excitonic condensate is traditionally described by a 
many-body wavefunction that encodes the coherence between electron 
and hole bands. Here we compare the mean-field ground state in the limit of 
dilute electron (hole) density with the ground state of a single 
electron-hole pair in double-layer system. As the interlayer distance $d$ 
increases, we find that the excitonic size, characterized by the width of the 
momentum-space wavefunction, also increases. By comparing the single-exciton 
wavefunction with the mean-field analysis, we determine the $d$-dependence 
of the ``diluteness'' of the exciton gas in a balanced double-layer system 
with given electron (or hole) density.
\end{abstract}

\pacs{}
\maketitle


\section{Introduction}
\label{sec:intro}
The Bose-Einstein condensation of excitons in double layer systems realized 
in semiconductor heterojunctions, where a 
macroscopic number of excitons occupy a single quantum state, has been a 
subject of intense research over the past decade.~\cite{bec,snokebook} An 
exciton is a metastable bound state of an electron and a hole. In a balanced 
electron-hole system at low electron (hole) densities $n_e=n_h=n_{2D}$, the 
distance between the excitons 
is much larger than the quantum exciton size and the excitons behave as 
weakly-interacting bosons.~\cite{snokebook,bl,keldysh} In a bulk 
semiconductor or a two-dimensional system, the internal state of the 
exciton is given by Hydrogenic wavefunctions~\cite{snokebook} that result 
from the Coulomb interaction $V_A(\br)=e^2/\epsilon r$. In each case, the 
wavefunction for the 
electron-hole separation decays exponentially with decay length $a_{ex}/2$ 
where $a_{ex}=\epsilon\hbar^2/e^2m_r$ is the quantum size of the exciton. Here 
$\epsilon\sim 10$ is the dielectric constant of the semiconductor 
heterojunction, $\br_e$ ($\br_h$) represents the electron (hole) position, 
$m_r^{-1}=m_e^{-1}+m_h^{-1}$ is the (reduced) mass of exciton, and $m_e$ 
($m_h$) is the electron (hole) band mass. Note that for a symmetric 
electron-hole system that we consider in this paper, $m_e=m_h=2m_r$, the 
quantum size of carriers is half the exciton size, 
$a_0=\epsilon\hbar^2/e^2m_e=a_{ex}/2$. Therefore the dimensionless distance 
between excitons for a given carrier density $n_{2D}$ is given by $r_s/2$ 
where $r_s$, defined by $\pi(r_s a_0)^2=1/n_{2D}$ is the dimensionless 
distance between the carriers. Since the quantum size $a_{ex}$ of the 
exciton is fixed by the semiconductor properties, it is possible to tune the 
interaction between excitons from weak ($r_s\gg 1$) to strong 
($r_s\sim 1$).~\cite{keldysh} Note that for a bulk or planar system, as 
opposed to a double-layer system, 
the inter-exciton interaction and the formation of exciton are both governed 
by the same Coulomb interaction $V_A(\br)$. 

In double-layer systems where electrons are carriers in the top layer and 
holes are the carriers in the bottom layer, the formation of an exciton is 
determined by the attractive interlayer Coulomb interaction 
$V_E(\br)=-e^2/\epsilon\sqrt{r^2+d^2}$ where $d$ is the distance between the 
two layers and $\br$ denotes the two-dimensional position vector. Since the 
attractive interaction is $d$ dependent, the quantum size of exciton is not 
necessarily $a_{ex}$ and, in fact, depends on the dimensionless ratio 
$d/a_0$. The interaction between the excitons, on the other hand, is also 
dependent on intralayer Coulomb repulsion $V_A(\br)$. Thus, in double-layer 
systems, the diluteness of excitons is a function of ($d/a_0,r_s$). In this 
paper, we quantitatively explore this issue. In the next section we present 
numerical solution to the single-exciton problem. We find that the exciton 
binding energy $E_b$ decreases as $d$ increases and concurrently the 
momentum-space ground state wavefunction sharpens. Thus, the quantum size of 
an exciton $a_{ex}(d)$ increases with $d$. In Sec.~\ref{sec:meanfield}, we 
complement the single-exciton results with mean-field analysis of the 
uniform excitonic condensate ground state for varying carrier density 
$n_{2D}$. By comparing the exciton wavefunction obtained from the Wannier 
approximation with that in Sec.~\ref{sec:se}, we obtain a quantitative 
criterion for the ``diluteness'' of an exciton gas. We conclude the paper 
with a remarks in Sec.~\ref{sec:disc}. 


\section{Single-exciton problem}
\label{sec:se}
Let us start with an electron and a hole in a double-layer system with $d=0$. 
The eigenstates of this problem are obtained by solving the equivalent 
problem of a particle with mass $m_r$ in a central attractive potential. Due 
to the rotational invariance in two dimensions~\cite{yang1991} and the 
existence of the conserved Runge-Lenz vector,~\cite{dittrich} the energy 
spectrum in the limit $d=0$ is dependent only on the principle quantum number 
$n\geq 1$, $E_n=-4E_0/(2n-1)^2$ 
where $E_0=e^2/\epsilon a_0$ is the energy scale associated with the problem. 
(Note that typical parameters $\epsilon\sim$ 10 and $a_0\sim$ 50\AA\, imply 
$E_0\sim$ 30 meV). The corresponding normalized ground-state wavefunction is 
given by $\psi_G(r)=\sqrt{8/(\pi a_{ex}^2)}\exp(-2r/a_{ex})=
\sqrt{2/(\pi a_0^2)}\exp(-r/a_0)$.~\cite{yang1991} 
When $d\neq 0$, since the electron-hole interaction is given by 
$V_E(\br)=-e^2/\epsilon\sqrt{r^2+d^2}$, the differential equation for the 
radial component of the excitonic wavefunction that cannot be analytically 
solved. Instead, we use the momentum-space Schr\"{o}dinger equation 
\begin{equation}
\label{eq:pspace}
\frac{\hbar^2 k^2}{2m_r}\psi_\alpha(\bk)+\int_{\bk'}V_E(|\bk-\bk'|)
\psi_{\alpha}(\bk')=E_\alpha\psi_\alpha(\bk)
\end{equation}
where $\bk$ is the two-dimensional wavevector, $\psi_\alpha(\bk)$ is the 
momentum-space eigenfunction with eigenvalue $E_\alpha$, and 
$V_E(\bq)=-V_A(\bq)e^{-qd}$ where $V_A(\bq)=2\pi e^2/\epsilon q$ is the 
Fourier transform of the intralayer Coulomb interaction in two dimensions. 
We focus on Eq.(\ref{eq:pspace}) 
projected onto the zero angular-momentum sector and obtain the eigenenergies 
and eigenfunctions by discretizing the integral equation and numerically 
diagonalizing the resulting matrix~\cite{karr2009}  
\begin{equation}
\label{eq:pspacediscrete}
H_{mn}=\frac{u_n^2}{2}\delta_{mn}+\frac{u_n\Delta u}{2\pi}\tilde{V}(u_m,u_n)= 
\frac{u_m}{u_n}H^{*}_{nm}.
\end{equation}
Here $\tilde{V}(u_m,u_n)$ is the angular-averaged dimensionless electron-hole 
interaction. Although the Hamiltonian (\ref{eq:pspacediscrete}) leads to 
bound and continuum states, since the excitonic internal states are only 
accessible at temperatures $T\geq E_0/k_B\sim 300$ K, in the following we only 
discuss the behavior of the ground state. 

\begin{figure}
\includegraphics[width=8.6cm]{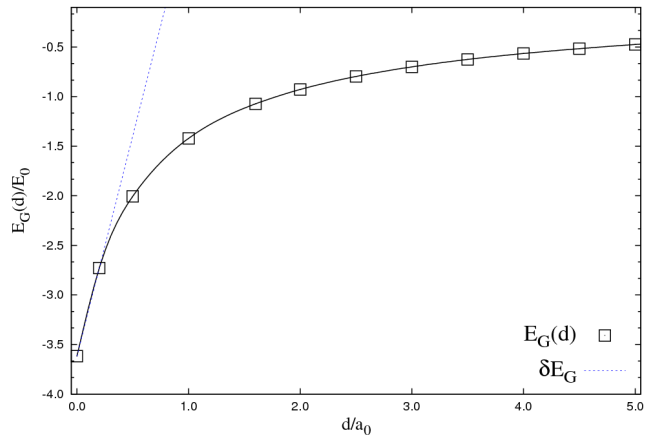}
\caption{\label{fig:ebound}(Color Online) 
Excitonic ground-state energy $E_G(d)$ as a function of interlayer distance 
$d$ obtained from single-particle Schr\"{o}dinger equation. The dotted line 
through $d=0$ shows that at small $d$ the change in the binding energy is 
linear, $\delta E_G=4E_0(d/a_0)$, as expected from first-order perturbation 
theory. The binding energy is strongly suppressed at large $d$.}
\end{figure}

Figure~\ref{fig:ebound} shows the numerically obtained ground-state energy of 
a single exciton as a function of interlayer distance $d$. At $d=0$, the 
numerical result deviates from the well-known analytical answer by 10\%; 
however, we have verified that this difference is solely due to 
discretization errors and can be systematically suppressed.~\cite{karr2009} 
When $d/a_0\ll 1$ first-order perturbation theory implies that the change in 
the ground-state energy is linear, $\delta E_G=E_G(d)-E_G(0)= 4E_0(d/a_0)$. 
At large $d$ the excitonic binding energy is strongly suppressed; for 
example, when $d/a_0=10$ it is reduced to 10\% of the binding energy at $d=0$. 

\begin{figure}
\includegraphics[width=8.6cm]{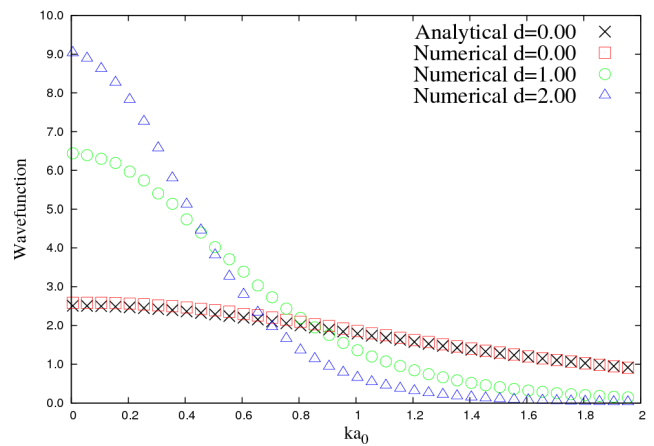}
\caption{\label{fig:psibound}(Color online) 
Ground state wave-function $\psi_G(k)$ for a single exciton as a function of 
interlayer distance $d$. The analytical (cross) and numerical (open square) 
solutions for $d=0$ are consistent with each other. Their momentum-space 
width indicates that the exciton size is $a_0$. As $d$ increases $\psi_G(k)$ 
sharpens and the size of the exciton, defined by the inverse-width of the 
momentum-space wavefunction, increases.}
\end{figure}

In Figure~\ref{fig:psibound}, we show the corresponding evolution of the 
ground-state wavefunction with increasing $d$. At $d=0$ the normalized 
wavefunction is given by $\psi_G(k)=\sqrt{8\pi a_0^2}/(1+k^2a_0^2)^{3/2}$ 
and is reproduced by our numerical calculations. As $d$ increases, we see that 
the momentum-space wavefunction sharpens and shows that the single exciton 
size $a_{ex}(d)$ increases with $d$. 

These results show that a single exciton in a double-layer system is 
increasingly weakly bound, and becomes larger as the interlayer distance $d$ 
increases. Therefore, although $r_s$ is a good measure of the diluteness of 
electrons or holes, it is not a good measure of diluteness for the excitonic 
gas. To quantify this observation, in the next section, we study the 
evolution of a uniform excitonic condensate state as a function of 
($d/a_0,r_s$).


\section{Mean-field analysis}
\label{sec:meanfield}
To explore the uniform excitonic condensate state, we start with a 
double-layer system with electrons in the top layer and holes in the bottom 
layer that is separated by distance $d$. The Hamiltonian for such a system is 
a sum of the kinetic energy for electrons and holes, as well as the 
intralayer Coulomb repulsion $V_{A}(\bq)$ and the interlayer Coulomb 
attraction $V_{E}(\bq)$. We use the standard mean-field 
approximation~\cite{negele,pbl,graphenechz} to obtain the mean-field 
Hamiltonian, 
\begin{eqnarray}
\label{eq:hmf} 
H &=& 
\sum_\bk (e^{\dagger}_\bk\, h_{-\bk})
\left(
\begin{array}{cc}
\xi_\bk\, & \Delta_\bk \\
\Delta^{*}_\bk\, & -\xi_\bk\\
\end{array}
\right)
\left(
\begin{array}{c}
e_\bk\\
h^{\dagger}_{-\bk}\\
\end{array}
\right)
\end{eqnarray}
where $e^{\dagger}_{\bk}$ ($h^{\dagger}_{\bk}$) is the creation operator for 
an electron (hole) with two-dimensional momentum $\hbar\bk$ in the top 
(bottom) layer, $\xi_\bk$ is renormalized electron (hole) dispersion that 
takes into account the interlayer capacitance and intralayer exchange energy, 
and $\Delta_\bk$ is the excitonic order parameter associated with the 
coherence between the electron and the hole bands in the two layers. We 
consider an isotropic excitonic-condensate order parameter and obtain the 
following self-consistent mean-field equations~\cite{pbl,graphenechz}
\begin{eqnarray}
\xi_k & = & \varepsilon_k+V_C-\mu-
\int_{\bk'}V_A(|\bk-\bk'|)\langle e^{\dagger}_{\bk'} e_{\bk'}\rangle
\label{eq:xi} 
\\
\Delta_k & = &\int_{\bk'}V_E(|\bk-\bk'|)\langle h_{-\bk'}e_{\bk}\rangle
\label{eq:delta}
\\
n_{2D} & = & \int_{\bk}\langle e^\dagger_{\bk}e_{\bk}\rangle=\frac{1}{2}
\int_{\bk}\left(1-\frac{\xi_k}{E_k}\right)
\label{eq:density}
\end{eqnarray}
where $\varepsilon_\bk=\hbar^2 k^2/2m_e=\hbar^2 k^2/2m_h$ denotes the 
electron (hole) band dispersion, $V_C=2\pi e^2 n_{2D}/\epsilon$ is the 
capacitive energy cost, and $\mu$ is the (electron and hole) chemical 
potential determined implicitly by Eq.(\ref{eq:density}). The self-consistent 
excitonic order parameter is determined by 
$\langle h_{-\bk}e_{\bk}\rangle=\Delta_k/2E_k$, and 
$\pm E_\bk=\pm\sqrt{\xi_\bk^2+\Delta_\bk^2}$ denote dispersion of the 
quasiparticle bands that result from Hamiltonian (\ref{eq:hmf}). We solve 
Eqs.(\ref{eq:xi})-(\ref{eq:density}) iteratively for a given $(d/a_0,r_s)$ to 
obtain the self-consistent order parameter $\Delta_k$ and quasiparticle energy 
dispersion $E_k$. To explore the dilute exciton limit, we recast 
Eq.(\ref{eq:delta}) in terms of $\Phi(p)=\Delta_p/E_p$, and note that for 
$\Delta_p\ll\xi_p$ Eq.(\ref{eq:delta}) reduces to the single-exciton 
Schr\"{o}dinger equation in momentum space, Eq.(\ref{eq:pspace}). This 
permits a quantitative comparison between the ground-state exciton 
wavefunction $\psi_G(p)$ and the Wannier-exciton wavefunction $\Phi(p)$. 

\begin{figure}
\includegraphics[width=8.6cm]{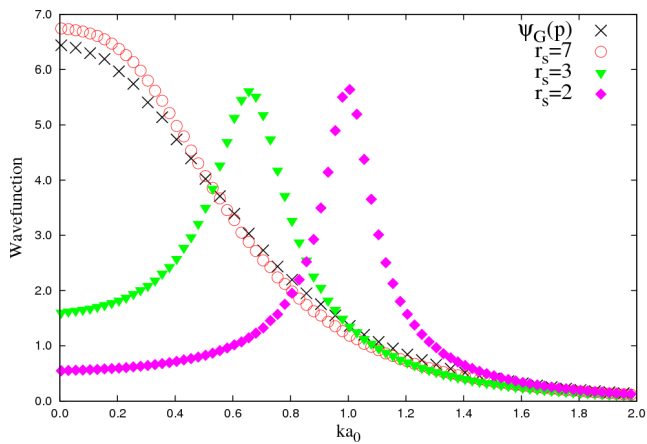}
\caption{\label{fig:d1compare}(Color online) Comparison of the Wannier 
wavefunctions $\Phi(p)=\Delta_p/E_p$ obtained from the mean-field solutions 
for $r_s=7$ (red circle), $r_s=3$ (green triangle), and $r_s=2$ (diamond), 
with the single-exciton wavefunction $\psi_G(p)$ (cross) for $d/a_0=1$. As 
$r_s$ increases the Wannier wavefunction approaches the single-exciton 
result.} 
\end{figure}
Figure~\ref{fig:d1compare} compares the Wannier wavefunction $\Phi(p)$ at 
$d/a_0=1$ for different values of $r_s$ with the single-exciton wavefunction 
$\psi_G(p)$. We see that $r_s$ increases the Wannier wavefunction approaches 
the single-particle result, as expected. Note that for small $r_s$, the 
Wannier exciton wavefunction $\Phi(p)$ is peaked at finite momentum because 
the excitonic order parameter $\Delta_p$ is maximum and $\xi_p$ is minimum at 
the Fermi momentum. However, as $r_s$ increases, for any given $d$, the peak 
in $\Delta_p$ shifts towards the origin and so does the maximum of the 
Wannier wavefunction. 

\begin{figure}
\includegraphics[width=8.6cm]{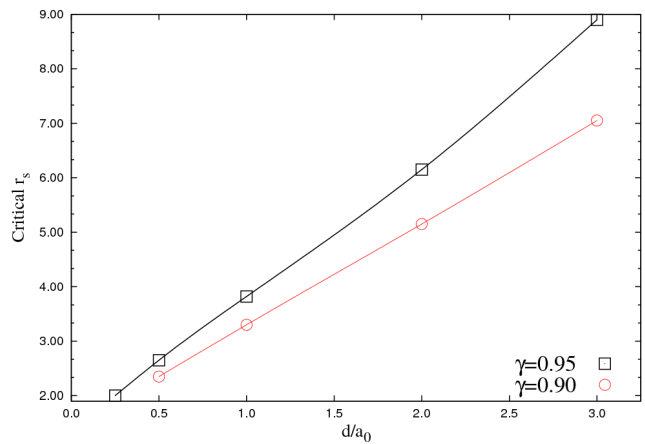}
\caption{\label{fig:overlap}(Color online) Interlayer-distance dependence of 
the critical $r_s$ value obtained using the constraints $\gamma=0.90$ 
(bottom) and $\gamma=0.95$ (top). $r_{sc}(d)$ provides a quantitative way to 
characterize the diluteness of an excitonic gas by comparing the Wannier 
wavefunction $\Phi(k)$ with the single-exciton solution $\phi_G(k)$.}  
\end{figure}

To quantify the proximity between the Wannier and the single-exciton 
approach, we consider the overlap $\gamma(d/a_0,r_s)$ between the two (real) 
wavefunctions
\begin{equation}
\label{eq:overlap}
\gamma(d/a_0,r_s)=\int\frac{d\bk}{(2\pi)^2}\Phi^{*}(k)\psi_G(k). 
\end{equation}
A high overlap value $\gamma(d/a_0,r_{sc})\sim 1$, allows us to define a 
critical value of $r_{sc}(d/a_0)$ such that for $r_s\geq r_{sc}$ the 
single-exciton result provides an excellent substitute for the mean-field 
analysis. Figure~\ref{fig:overlap} shows the critical $r_{sc}(d)$ obtained 
using $\gamma=0.90$ and $\gamma=0.95$. We see that for typical values of $d$ 
the critical $r_{sc}$ scales linearly with $d/a_0$. It implies, for example, 
that approaching the dilute-limit at $d/a_0=3$ will require reducing the 
carrier density by a factor of 5 from the corresponding value for the dilute 
limit at $d/a_0=1$.    


\section{Discussion}
\label{sec:disc}

The subject of excitonic condensation in double-layer systems has been 
extensively explored in the literature; the properties of a single exciton in 
a double-layer system, however, have not been. In this paper, we have 
obtained the ground-state wavefunction and the ground-state energy for a 
single exciton as a function of interlayer distance $d$. By comparing our 
results with those from a mean-field analysis of the uniform excitonic 
condensate, we have obtained the critical value $r_{sc}(d)$ that is used 
to determine when the exciton gas for a given interlayer distance $d$ is 
``dilute''. 

Our analysis provides a quantitative picture of a single exciton in 
double-layer system with $d\neq 0$ where analytical solution for the 
excitonic wave-functions is not possible. It shows that as $d$ increases, due 
to the weakened electron-hole Coulomb interaction, the exciton size 
$a_{ex}(d)$ increases. 

We note that the Wannier approximation for excitonic 
wavefunction is based on a mean-field analysis that usually 
over-estimates~\cite{negele} the excitonic order parameter $\Delta_k$. In 
particular, in double layer systems, it is known that the uniform condensate 
becomes unstable~\cite{palo} when $d$ is larger than a critical layer 
separation $d_c$.  Thus, when fluctuations around the mean-field 
state are taken into account, the critical value of $r_s$ for a given 
$d\sim d_c$ will change substantially. Our mean-field analysis is based on a 
uniform excitonic condensate state. Due to dipolar repulsion between 
excitons, the uniform state is unstable towards formation of a crystalline 
excitonic condensate~\cite{wignerss} in the region 
$\sqrt{r_s}\ll d/a_0\ll r_s$. Since our calculations lie outside this 
parameter range, we have focused only on the uniform state; the question of 
a ``dilute'' exciton limit in a crystalline excitonic condensate, however, 
remains open. 




\end{document}